\documentclass[amsmath,prb,twocolumn,showpacs,superscriptaddress]{revtex4}

\usepackage{graphicx}
\usepackage{boxedminipage}

\begin{document}

\title{Resonant inelastic tunneling in molecular junctions}

\date{\today}

\author{Michael Galperin}
\affiliation{Department of Chemistry and Nanotechnology Center, Northwestern University, Evanston IL 60208, U.S.A.}
\author{Abraham Nitzan}
\affiliation{School of Chemistry, Tel Aviv University, Tel Aviv 69978, Israel}
\author{Mark A. Ratner}
\affiliation{Department of Chemistry and Nanotechnology Center, Northwestern University, Evanston IL 60208, U.S.A.}

\begin{abstract}
Within a phonon-assisted resonance level model we develop a self-consistent
procedure for calculating electron transport
currents in molecular junctions with intermediate to strong 
electron-phonon interaction. The scheme takes into account the mutual influence 
of the electron and phonon subsystems. It is based on the $2^{nd}$ order 
cumulant expansion, used to express the correlation function of the phonon 
shift generator in terms of the phonon momentum Green function. 
Equation of motion (EOM) method is used to obtain
an approximate analog of the Dyson equation for the electron and phonon 
Green functions in the case of many-particle operators present in the 
Hamiltonian.
To zero-order it is similar in particular 
cases (empty or filled bridge level) to approaches proposed earlier. 
The importance of self-consistency in resonance tunneling situations
(partially filled bridge level) is stressed. 
We confirm, even for strong vibronic coupling, 
a previous suggestion concerning the absence
of phonon sidebands in the current vs. gate voltage plot when
the source-drain voltage is small~\cite{Mitra}.
\end{abstract}


\maketitle

\section{\label{intro}Introduction}
Molecular electronics is an active area of research with possible
technological interest in
supplementing currently available $Si$ based electronics via further 
miniaturization of electronic devices~\cite{MEB,AN}. 
Experiments on conduction in 
molecular junctions are becoming more common~\cite{Reed,Bowler}.
Early experiments focused on
the absolute conduction and on trends such as dependence on wire length,
molecular structure, and temperature. An intriguing issue is the role
played by molecular nuclear motions. Vibronic coupling may lead to rotations,
conformational changes, atomic rearrangements, and chemical reactions
induced by the electronic current~\cite{Park,Ho_conf}.
It is directly relevant to the
junction heating problem and can manifest itself in polaron-type localization
effects and inelastic signals in current-voltage spectra.
\cite{Ho_IETS,Zhitenev,Weig,Natelson,Ruitenbeck,Reed_IETS,Kushmerick,SelzerMayer}
While weak inelastic signatures are observed in measurements of
conductance with strong electronic coupling between the 
molecule and leads, stronger vibronic peaks
with pronounced phonon sidebands (we use the term 
phonon to describe any vibrational excitation) may be observed
when this coupling is weak. A full Franck-Condon envelope has been
observed with weak molecule-electrode interactions at both termini~\cite{Ho_FC}.

Studies of electron-phonon interaction have a long history~\cite{Mahan},
however
new points for consideration arise in biased current carrying junctions.
The interpretation of electronic transport in molecular 
junctions has until recently has been made largely in the context of 
multi-channel scattering problems~\cite{DiVentra,Todorov,Seideman,NessFisher,
BoncaTrugman,Troisi}, which disregards the 
influence of the contact population
(manifestations of the Pauli principle blocking of scattering channels, change
of electronic structure, etc.) on inelastic process.
Such approaches also disregard the influence of the electronic subsystem
on the phonon dynamics. A systematic framework describing transport phenomena
of many-particle systems which can take these effects into account
can be developed based on the non-equilibrium Green's function (NEGF)
formulation\cite{Keldysh,DattaBook,HaugJauho}.

Phonon-assisted electron transport in molecular systems can be 
classified by the relative time and energy scales of the processes involved. 
The electron lifetime in the junction should be compared to the relevant 
vibrational frequency\cite{Burin}, 
while the strength of the electron-phonon coupling should be 
judged relative to electronic matrix elements (coupling to contacts and/or 
between isolated parts of the molecular system). It is useful to
consider separately the limits of weak and strong electron-phonon coupling.  
The first corresponds to non-resonant phonon-assisted electron tunneling
mostly encountered in experiments on inelastic electron tunneling spectroscopy
(IETS)~\cite{Ho_IETS,Reed_IETS,Kushmerick,SelzerMayer,Bayman}.
With the development and advances in scanning tunneling microscopy
(STM) and scanning tunneling spectroscopy (STS), IETS has proven invaluable as
a tool for identifying and characterizing molecular species within the
conduction region. The use of Migdal-Eliashberg theory\cite{ME,FetterWalecka} 
is justified in this case, whereupon the lowest non-vanishing (second) order 
perturbation in electron-phonon
coupling on the Keldysh contour leads to the Born approximation
(BA) for electron dynamics. This approach using BA or its self-consistent
(in electron Green function) flavor was used in several
theoretical 
studies~\cite{Ehrenreich,Hershfield,Datta,Ueba,Pecchia,Mitra,Yamamoto}.
In a recent publication~\cite{IETS_SCBA} we used an advanced version of this
scheme, self-consistent Born approximation (SCBA) for both electron and phonon
Green functions, to describe features (peaks, dips, line shape, and line width)
of the IETS signal, $d^2I/d\Phi^2$, as a function of the applied voltage $\Phi$.
Sometimes SCBA is used also in the resonant tunneling 
regime~\cite{Mitra,Keller}.
This usage is valid only if electron-phonon interaction is weak, so that
no essential inelastic features (e.g. polaron formation) can be studied 
in this case.

The other limit is realized in cases of resonant tunneling, which
is characterized by longer electron lifetime in the
junction (though it still may be short relative to the characteristic phonon 
frequency) and stronger effective electron-phonon coupling. 
The perturbative treatment breaks down in this case which
may result in formation of a polaron in the junction. 
Signatures of resonant tunneling driven by an intermediate 
molecular ion appear as peaks in the first derivative $dI/d\Phi$ and may show 
phonon subbands~\cite{mceuen,ralph,natelson}. 
Several theoretical studies of this situation in tunneling junctions are 
available~\cite{NessFisher,BoncaTrugman,Wingreen,Lundin,Balatsky,Bratkovsky}.
Most of them~\cite{NessFisher,BoncaTrugman,Wingreen} are based on scattering 
theory consideration, others~\cite{Lundin,Balatsky,Bratkovsky} are based
partially on the NEGF methodology. However these works disregard 
the Fermi population in the leads as mentioned above. 
Another approach, the non-equilibrium linked cluster expansion (NLCE), 
is based on generalization of the linked cluster expansion
to nonequilibrium situations~\cite{Kral}.
This approach takes the contact population into account, but appears
to be unstable for diagrammatic expansion beyond the lowest order. 
Note also that in all the cases mentioned above the phonon subsystem is 
assumed to remain in thermal equilibrium throughout the process.
The rate equations approach often used in the literature for the case of 
weak coupling
to the leads~\cite{Mitra,BraigFlensberg,vonOppen} is essentially a 
quasiclassical treatment, having an assumption that the tunneling rate is 
much smaller than decoherence rates on the molecular subsystem. 
While the approach is 
useful in describing e.g. $C_{60}$ center of mass motion as is done in 
Ref.~\onlinecite{BraigFlensberg}, the intramolecular vibrations we are interested 
in here may have a much longer lifetime, thus making the rate equations
approach inappropriate. An attempt to generalize single particle 
approximation of Refs.~\onlinecite{Lundin,Balatsky,Bratkovsky} was presented in
Ref.~\onlinecite{Flensberg}, where an EOM method was used. 
Another interesting approach uses numerical renormalization group
methodology to study inelastic effects in conductance in the
linear response regime~\cite{Cornaglia}.

In this paper we propose an approximate scheme for treating electronic
transport in cases involving intermediate to strong electron-phonon coupling 
in tunnel junctions. We employ the Keldysh contour based description,
which treats both electron and phonon degrees 
of freedom in a self-consistent manner. 
This approach is close in spirit to NLCE~\cite{Kral} in using a cumulant 
expansion (when finding an approximate
expression for phonon shift operators correlation function  in terms of
phonon Green functions). However unlike NLCE the proposed scheme is stable
and self-consistent, i.e. the influence of tunneling current on 
the phonon subsystem is taken into account. 
It reduces to the scattering theory results in the limit where
the molecular bridge energies are far above the Fermi energy of the leads
and provides a scheme for analyzing the effect of the electronic current
on the energetics in the vibrational space in resonance
tunneling situations (the issue will be discussed elsewhere).
Derivation of equations is based on EOM method, which makes our
approach similar to Ref.~\onlinecite{Flensberg}. However we go beyond it
in taking into account renormalization of phonon subsystem due to 
coupling to tunneling electron. 
Thermal relaxation of the molecular phonons, not taken into account
in~\cite{Flensberg}, is also introduced in our consideration.

In the next section we introduce the model and describe the approximations made.
Section~\ref{sc_scheme} presents the procedure of our 
self-consistent calculation.
In section~\ref{numerics} we report numerical results 
and compare them to results obtained within other approaches.
Section~\ref{conclude} concludes.

\section{\label{model}Model}
We consider a simple resonant-level model with the electronic level $|0>$ 
coupled to two electrodes left ($L$) and right ($R$) (each a free electron 
reservoir at its own equilibrium).  The electron on the resonant level 
(electronic energy $\varepsilon_0$) is linearly coupled to a single vibrational 
mode (phonon) with frequency $\omega_0$, henceforth referred to as the 
``primary phonon''.
The latter is coupled to a phonon bath represented as a set of independent 
harmonic oscillators. The system Hamiltonian is
(here and below we use $\hbar=1$ and $e=1$)
\begin{align}
 \label{H}
 &\hat H = \varepsilon_0\hat c^\dagger\hat c +
 \sum_{k\in\{L,R\}} \varepsilon_k \hat c_k^\dagger\hat c_k +
 \sum_{k\in\{L,R\}} \left(V_k\hat c_k^\dagger\hat c + \mbox{h.c.}\right) 
 \nonumber \\ &+
 \omega_0\hat a^\dagger\hat a + 
 \sum_\beta\omega_\beta\hat b^\dagger_\beta\hat b_\beta +
 M_a\hat Q_a\hat c^\dagger\hat c +
 \sum_\beta U_\beta\hat Q_a\hat Q_\beta
\end{align}
where $\hat c^\dagger$ ($\hat c$) are creation (destruction) operators
for electrons on the bridge level, $\hat c_k^\dagger$ ($\hat c_k$) are 
corresponding operators for electronic states in the contacts, 
$\hat a^\dagger$ ($\hat a$)
are creation (destruction) operators for the primary phonon, and
$\hat b_\beta^\dagger$ ($\hat b_\beta$) are the corresponding operators for
phonon states in the thermal (phonon) bath. 
$\hat Q_a$ and $\hat Q_\beta$ are phonon displacement operators
\begin{equation}
 \label{Q}
 \hat Q_a = \hat a + \hat a^\dagger \qquad
 \hat Q_\beta = \hat b_\beta + \hat b_\beta^\dagger
\end{equation}
The energy parameters $M_a$ and $U_\beta$ correspond to the vibronic
and the vibrational coupling respectively.
For future reference we also introduce the phonon momentum operators
\begin{equation}
 \label{P}
 \hat P_a = -i\left(\hat a - \hat a^\dagger\right) \qquad
 \hat P_\beta = -i \left(\hat b_\beta - \hat b_\beta^\dagger\right)
\end{equation}
In what follows we will refer to the phonon mode $a$ that is directly
coupled to the electronic system as the ``primary phonon''.

Following previous work on strong electron-phonon 
interaction~\cite{Lundin,Balatsky,Bratkovsky} we start by applying
a small polaron (canonical or Lang-Firsov) 
transformation~\cite{Mahan,LangFirsov} to the Hamiltonian (\ref{H})
\begin{equation}
 \label{small_polaron}
 \hat{\bar H} = e^{\hat S_a}\hat H e^{-\hat S_a}
\end{equation}
with
\begin{equation}
 \label{Sa}
 \hat S_a = \frac{M_a}{\omega_0}\left(\hat a^\dagger-\hat a\right)
 \hat c^\dagger\hat c
\end{equation}
Under the additional approximation of neglecting
the effect of this transformation on the coupling of 
the primary phonon to the thermal phonon bath this leads to
%(in the spirit of the non-crossing approximation (NCA)~\cite{Bickers})
\begin{eqnarray}
 \label{barH}
 \hat{\bar H} &=& \bar\varepsilon_0\hat c^\dagger\hat c +
 \sum_{k\in\{L,R\}} \varepsilon_k \hat c_k^\dagger\hat c_k +
 \sum_{k\in\{L,R\}} \left(V_k\hat c_k^\dagger\hat c\hat X_a+\mbox{h.c.}\right)
 \nonumber \\ &+&
 \omega_0\hat a^\dagger\hat a +
 \sum_\beta\omega_\beta\hat b^\dagger_\beta\hat b_\beta +
 \sum_\beta U_\beta\hat Q_a\hat Q_\beta
\end{eqnarray}
where 
\begin{equation}
 \label{bare0}
 \bar\varepsilon_0 = \varepsilon_0 - \Delta \qquad
 \Delta \approx \frac{M_a^2}{\omega_0};
\end{equation}
$\Delta$ is the electron level shift due to coupling to the primary phonon and
\begin{equation}
 \label{Xa}
 \hat X_a = \exp\left[i\lambda_a\hat P_a\right] \qquad
 \lambda_a = \frac{M_a}{\omega_0}
\end{equation}
is primary phonon shift generator. The Hamiltonian (\ref{barH})
is characterized by the absence of direct electron-phonon coupling 
present in (\ref{H}).
This is replaced by renormalization of coupling to the contacts.
Note that the same result can be obtained by repeatedly applying
the transformation analogous to (\ref{Sa}) in the case
of weak coupling between primary phonon and thermal bath
and neglecting renormalization due to thermal bath phonons 
(see Appendix~\ref{appA}).

The Hamiltonian (\ref{barH}) is our starting point for the calculation of the
steady-state current across the junction, using the
NEGF expression derived in Refs.~\onlinecite{HaugJauho,current} 
\begin{equation}
 \label{current}
 I_K = \frac{e}{\hbar}\int\frac{dE}{2\pi}
 \left[\Sigma_K^{<}(E)\, G^{>}(E) - \Sigma_K^{>}(E)\, G^{<}(E)\right]
\end{equation}
Here $\Sigma_K^{<,>}$ are lesser/greater projections of the self-energy
due to coupling to the contact $K$ ($K=L,R$)
\begin{eqnarray}
 \label{SEKlt}
 \Sigma_K^{<}(E) &=& i f_K(E) \Gamma_K(E) \\
 \label{SEKgt}
 \Sigma_K^{>}(E) &=& -i [1-f_K(E)] \Gamma_K(E)
\end{eqnarray}
with $f_K(E)$ the Fermi distribution in the contact $K$ and 
\begin{equation}
 \label{GammaK}
 \Gamma_K(E) = 2\pi \sum_{k\in K} |V_k|^2 \delta(E-\varepsilon_k)
\end{equation}

The lesser and greater Green functions in (\ref{current}) are 
Fourier transforms to energy space of projections onto 
the real time axis of the electron Green function on the Keldysh contour 
\begin{align}
 \label{GFKeldysh}
 G(\tau_1,\tau_2) &= -i<T_c \hat c(\tau_1)\hat c^\dagger(\tau_2)>_H
 \nonumber \\
 &= -i<T_c \hat c(\tau_1)\hat X_a(\tau_1)\,
          \hat c^\dagger(\tau_2)\hat X_a^\dagger(\tau_2)>_{\bar H}
\end{align}
where the subscripts $H$ and $\bar H$ indicate which Hamiltonian, 
(\ref{H}) or (\ref{barH}) respectively,
determines evolution of the system, and $T_c$ is the contour ordering
operator. We use the second form and make the
usual approximation of decoupling electron and phonon dynamics
\begin{equation}
 \label{appGFKeldysh}
 G(\tau_1,\tau_2) \approx G_c(\tau_1,\tau_2)\, K(\tau_1,\tau_2)
\end{equation}
where
\begin{eqnarray}
 \label{Gc}
 G_c(\tau_1,\tau_2) &=& -i<T_c \hat c(\tau_1)\hat c^\dagger(\tau_2)>_{\bar H}
 \\
 \label{K}
 K(\tau_1,\tau_2) &=& <T_c \hat X_a(\tau_1)\hat X_a^\dagger(\tau_2)>_{\bar H}
\end{eqnarray}
Below we drop the subscript $\bar H$ keeping in mind that Hamiltonian 
(\ref{barH}) is the one that determines the time evolution of the system.
Previous studies (see e.g. \cite{Lundin,Balatsky,Bratkovsky}) 
stopped here and approximated the functions $G_c$ and $K$ 
by the electron Green function in the absence
of coupling to phonons and by the equilibrium 
correlation function of the phonon shift generator $\hat X_a$, respectively.
Here we go a step further by 'dressing' these terms
in the spirit of standard diagrammatic technique~\cite{Mahan}. 
This will yield a self-consistent approach for the intermediate to strong 
electron-phonon interaction, the strong coupling analog of 
the self-consistent Born approximation (SCBA) used in the weak coupling 
limit~\cite{IETS_SCBA}.

We start by expressing the shift generator correlation function $K$
in terms of the phonon Green function. One can show (see Appendix~\ref{appB})
that in the second order cumulant expansion this connection is
\begin{equation}
 \label{XXKeldysh}
 K(\tau_1,\tau_2) =
 \exp\left\{\lambda_a^2\left[i D_{P_aP_a}(\tau_1,\tau_2)
                -<\hat P_a^2>\right]\right\}
\end{equation}
where $\hat P_a$ is the primary phonon momentum operator defined in (\ref{P})
and 
\begin{equation}
 \label{D}
 D_{P_aP_a}(\tau_1,\tau_2) = -i<T_c \hat P_a(\tau_1)\hat P_a(\tau_2)>
\end{equation}
is the phonon momentum Green function.
$<\hat P_a^2>=i D_{P_aP_a}^{<,>}(t,t)$ is time independent in steady-state. 

Next we derive self-consistent equations for the electron Green function
$G_c(\tau_1,\tau_2)$, Eq.(\ref{Gc}),
and the phonon momentum Green function $D_{P_aP_a}(\tau_1,\tau_2)$, 
Eq.~(\ref{D}). Since we consider situations where the 
electron-phonon coupling is strong relative to the coupling between 
the bridge electronic level and the contacts it is
reasonable to look for an expression of second order in $V_k$. 
We use an equation of motion (EOM) method to obtain expressions for 
these Green functions. Since the Hamiltonian (\ref{barH}) contains
the exponential operator $\hat X$, Wick's theorem is not applicable
so we cannot write the Dyson equations to express these Green functions
in terms of the corresponding self-energies. It is nevertheless
possible to obtain the following approximate coupled integral equations for 
these Green functions (see Appendices~\ref{appC} and \ref{appD})
\begin{align}
 \label{DKeldysh}
 &D_{P_aP_a}(\tau,\tau') = D_{P_aP_a}^{(0)}(\tau,\tau')
 \\
 &+ \int_c d\tau_1 \int_c d\tau_2\, D_{P_aP_a}^{(0)}(\tau,\tau_1)\,
 \Pi_{P_aP_a}(\tau_1,\tau_2)\, D_{P_aP_a}^{(0)}(\tau_2,\tau') 
 \nonumber \\
 \label{GcKeldysh}
 &G_c(\tau,\tau') = G_c^{(0)}(\tau,\tau')
 \\
 &+ \sum_{K=\{L,R\}}\int_c d\tau_1 \int_c d\tau_2\, G_c^{(0)}(\tau,\tau_1)\,
 \Sigma_{c,K}(\tau_1,\tau_2)\, G_c^{(0)}(\tau_2,\tau')
 \nonumber
\end{align}
where the functions $\Pi_{P_aP_a}$ and $\Sigma_{c,K}$ are given by
\begin{align}
 \label{DSEKeldysh}
 &\Pi_{P_aP_a}(\tau_1,\tau_2) = \sum_\beta |U_\beta|^2 
 D_{P_\beta P_\beta}(\tau_1,\tau_2) 
 - i\lambda_a^2\sum_{k\in\{L,R\}}|V_k|^2
 \nonumber \\
 &\times\left[
     g_k(\tau_2,\tau_1)G_c(\tau_1,\tau_2)
     <T_c \hat X_a(\tau_1)\hat X_a^\dagger(\tau_2)>
  +  (\tau_1\leftrightarrow\tau_2)\right]
 \\
 \label{GcSEKeldysh}
 &\Sigma_{c,K}(\tau_1,\tau_2) = \sum_{k\in K} |V_k|^2 g_k(\tau_1,\tau_2)
 <T_c \hat X_a(\tau_2)\hat X_a^\dagger(\tau_1)>
\end{align} 
Here $K=L,R$ and $g_k$ is the free electron Green function for state $k$ 
in the contacts.
Note that $\Pi_{P_aP_a}$ and $\Sigma_{c,K}$ play here the same role as
self-energies in the Dyson equation. 
Dyson-like equations of this kind are often used as approximations 
when Wick's theorem doesn't work (see e.g. Ref.\onlinecite{Silbey}).
Eqs.~(\ref{XXKeldysh}), (\ref{DKeldysh}) and (\ref{GcKeldysh}) can be solved 
self-consistently as described in the next Section.

\section{\label{sc_scheme}Self-consistent calculation scheme}
In what follows we use the term ``self-energy'' for the functions
$\Pi_{P_aP_a}$ and $\Sigma_{c,K}$, in analogy to the corresponding
functions that appear in true Dyson equations.
Eq.~(\ref{XXKeldysh}), (\ref{DKeldysh}) and (\ref{GcKeldysh}) 
provide a self-consistent way to solve for the phonon and electron
Green functions, $D_{P_aP_a}$ and $G_c$ and the corresponding self-energies 
$\Pi_{P_aP_a}$ and $\Sigma_{c,K}$.

Since the connection between the correlation function of the shift generators 
and the phonon
Green function, Eq.(\ref{XXKeldysh}), is exponential, it is easier to express
projections of the former function in terms of the latter one in the time 
domain. At the same time,
zero-order (no coupling between electron and phonon) expressions for the
lesser and greater projections of Green functions $G_c$ and $D_{P_aP_a}$
are easier to write down in the energy domain. As a result, we work in both
domains and implement fast Fourier transform (FFT) to transform between them. 
The ensuing self-consistent calculation scheme consists of the following 
steps
\begin{enumerate} 
\item We start with the zero-order retarded phonon and electron
Green functions in the energy domain
\begin{align}
 \label{D0rE}
 D_{P_aP_a}^{(0),r}(E) &= \left[E-\omega_0+i\frac{\gamma_{ph}}{2}\right]^{-1}
 \nonumber \\
    &- \left[E+\omega_0+i\frac{\gamma_{ph}}{2}\right]^{-1}
 \\
 \label{G0crE}
 G_c^{(0),r}(E) &= \left[E-\bar\varepsilon_0-\Sigma_c^{(0),r}(E)\right]^{-1}
\end{align}
where we use the wide-band limit (see Ref.~\onlinecite{IETS_SCBA} for discussion) 
for the phonon retarded self-energy due to coupling to thermal bath 
(the retarded projection of the first term on the right in 
Eq.(\ref{DSEKeldysh}))
\begin{equation}
 \label{gammaph}
 \gamma_{ph} = 2\pi\sum_\beta |U_\beta|^2\delta(E-\omega_\beta)
\end{equation}
and where the retarded electron self-energy due to coupling to the contacts
is taken in the form
\begin{eqnarray}
 \label{SE0crE}
 \Sigma_c^{(0),r}(E) &=& \Sigma_{c,L}^{(0),r}(E) + \Sigma_{c,R}^{(0),r}(E)
 \\
 \label{SE0crKE}
 \Sigma_{c,K}^{(0),r}(E) &=& \frac{1}{2}
 \frac{\Gamma_K^{(0)}W_K^{(0)}}{E-E_K^{(0)}+iW_K^{(0)}}
 \quad (K=L,R)
\end{eqnarray} 
with $E_K^{(0)}$ and $W_K^{(0)}$ being the center and half width of the band
respectively and $\Gamma_K^{(0)}$ is the escape rate
to contact $K$ in the electron wide band limit ($W_K^{(0)}\to\infty$ 
relative to all other energy parameters of the junction).

\item The lesser and greater projections of the phonon and electron Green 
functions are obtained using the Keldysh equation\cite{HaugJauho}
\begin{eqnarray}
 \label{DltgtE}
 D_{P_aP_a}^{<,>}(E) &=& |D_{P_aP_a}^{r}(E)|^2 \Pi_{P_aP_a}^{<,>}(E)
 \\
 \label{GcltgtE}
 G_c^{<,>}(E) &=& |G_c^{r}(E)|^2 \Sigma_c^{<,>}(E)
\end{eqnarray}
In the first iteration step we use the zero-order retarded Green functions 
(\ref{D0rE}) and (\ref{G0crE}) with the phonon self-energy due to coupling
to the thermal bath in place of $\Pi^{<,>}(E)$
\begin{align}
 \label{PiphltgtE}
 &|D_{P_aP_a}^{(0),r}(E)|^2\Pi_{P_aP_a,ph}^{<,>}(E) = 
 \nonumber \\
 &\quad -iN(E)\frac{\gamma_{ph}}{(E\mp\omega_0)^2+(\gamma_{ph}/2)^2}
 \\ 
 &\quad -i[1+N(E)]\frac{\gamma_{ph}}{(E\pm\omega_0)^2+(\gamma_{ph}/2)^2}
 \nonumber
\end{align}
where $N(E)$ is the Bose distribution of the thermal phonon bath.
In Eq.(\ref{PiphltgtE}) upper (lower) sign corresponding to lesser (greater) 
projection. Similarly the zero-order (in the vibronic coupling)
electron self-energy (sum of contributions due to left, $L$, and right, 
$R$, contacts) is used in place 
of $\Sigma_c^{<,>}(E)$
\begin{align}
 \label{SE0cltKE}
 \Sigma_{c,K}^{(0),<}(E) &= if_K(E)\Gamma_K(E) 
 \\
 \label{SE0cgtKE}
 \Sigma_{c,K}^{(0),<}(E) &= -i[1-f_K(E)]\Gamma_K(E)
 \\
 \label{GammaKE}
 \Gamma_K(E) &= -2\mbox{Im}[\Sigma_{c,K}^{(0),r}(E)] 
 \nonumber \\
            &= \frac{\Gamma_K^{(0)}(W_K^{(0)})^2}{(E-E_K^{(0)})^2+(W_K^{(0)})^2}
\end{align}
where $K=L,R$ and $f_K(E)$ is the Fermi distribution in contact $K$.
The lesser and greater projections are then transformed to the time domain.

\item\label{sc_start}
Utilizing Langreth rules~\cite{HaugJauho,Langreth} and using 
$D_{P_aP_a}^{<,>}(t)$ in projections of (\ref{XXKeldysh}) 
one can get lesser and greater correlation functions for the shift generator 
operators~\cite{Dltgtcomment} 
\begin{eqnarray}
 \label{XXlt}
 <\hat X_a^\dagger(0)\hat X_a(t)> &=& 
% \exp\left[i\lambda_a^2\left(D_{P_aP_a}^{<}(t)-D_{P_aP_a}^{<}(t=0)\right)\right]
 e^{i\lambda_a^2\left[D_{P_aP_a}^{<}(t)-D_{P_aP_a}^{<}(t=0)\right]}
 \\
 \label{XXgt}
 <\hat X_a(t)\hat X_a^\dagger(0)> &=&
% \exp\left[i\lambda_a^2\left(D_{P_aP_a}^{>}(t)-D_{P_aP_a}^{>}(t=0)\right)\right]
 e^{i\lambda_a^2\left[D_{P_aP_a}^{>}(t)-D_{P_aP_a}^{>}(t=0)\right]}
\end{eqnarray}

\item Using $\Sigma_c^{<,>}(t)$, $G_c^{<,>}(t)$, 
$<\hat X_a^\dagger(0)\hat X_a(t)>$ and $<\hat X_a(t)\hat X_a^\dagger(0)>$
in projections of the second term on the right in Eq.(\ref{DSEKeldysh})
yields the lesser and greater phonon self-energies due to coupling to the 
electron in the time domain
\begin{align}
 &\Pi_{P_aP_a,el}^{<}(t) = i\lambda_a^2<\hat X_a^\dagger(0)\hat X_a(t)>
 \\ &\quad\times
 \left\{\left[\Sigma_{c}^{(0),>}(t)\right]^{*}G_c^{<}(t)
 + \Sigma_{c}^{(0),<}(t)\left[G_c^{>}(t)\right]^{*}\right\}
 \nonumber \\
 &\Pi_{P_aP_a,el}^{>}(t) = i\lambda_a^2<\hat X_a(t)\hat X_a^\dagger(0)>
 \\ &\quad\times
 \left\{\Sigma_{c}^{(0),>}(t)\left[G_c^{<}(t)\right]^{*}
 + \left[\Sigma_{c}^{(0),<}(t)\right]^{*}G_c^{>}(t)\right\}
 \nonumber
\end{align} 
Similarly, projections of Eq.(\ref{GcSEKeldysh}) lead to lesser and greater 
electron self-energies in the time domain
\begin{eqnarray}
 \Sigma_c^{<}(t) &=& \Sigma_{c}^{(0),<}(t)<\hat X_a^\dagger(0)\hat X_a(t)>
 \\
 \Sigma_c^{>}(t) &=& \Sigma_{c}^{(0),>}(t)<\hat X_a(t)\hat X_a^\dagger(0)>
\end{eqnarray}
The retarded self-energies in time domain can be obtained from the lesser and 
greater counterparts in the usual way 
$\Sigma_c^r(t) = \theta(t)\left[\Sigma_c^{>}(t)-\Sigma_c^{<}(t)\right]$
In practice we use the time domain analog of the Lehmann 
representation~\cite{Mahan}
$\Sigma_c^r(t) = e^{-\delta t}\left[\Sigma_c^{>}(t)-\Sigma_c^{<}(t)\right]$
with $\delta\to 0$ to suppress negative time contributions on the FFT grid.
The retarded phonon self-energy is obtained in the same way.
Thus calculated self-energies are transformed to the energy domain.

\item\label{sc_end}
The self-energies obtained in the previous step are used to update 
the Green functions (retarded, lesser and greater; phonon and electron) 
in the energy domain. Retarded Green functions are 
\begin{eqnarray}
 \label{DrE}
 D_{P_aP_a}^r(E) &=& 
 \left[1/D_{P_aP_a}^{(0),r}(E)-\Pi_{P_aP_a,el}^r(E)\right]^{-1}
 \\
 \label{GcrE}
 G_c^r(E) &=& \left[E-\bar\varepsilon_0-\Sigma_c^r(E)\right]^{-1}
\end{eqnarray}  
The lesser and greater Green functions are then obtained from the Keldysh 
equation,
Eqs.~(\ref{DltgtE}) and (\ref{GcltgtE}). Note that the phonon self-energy
there contains contributions due to both coupling to the thermal bath 
and to the electron,
while the electron self-energy is a sum of contributions from the two contacts
dressed by the electron-phonon interaction.
The Green functions are then transformed to time domain.

\item The updated Green functions of the previous step are used
in step~\ref{sc_start}, closing the self-consistent loop.
Steps~\ref{sc_start}-\ref{sc_end} are repeated until convergence is achieved.
As a test we use electron population on the level
\begin{equation}
 n_0 = \mbox{Im}\left[G_c^{<}(t=0)\right]
\end{equation} 
Convergence is achieved when the absolute change of the level population 
in two subsequent iterations is less than a predefined tolerance.
Calculational grid is chosen fine enough to support the smallest energies 
and times involved and big enough to span over the relevant area. 
\end{enumerate}
The numerical steps described above require repeated integrations in
time and energy spaces. The numerical grids used for these integrations
are chosen so as to yield converged numerical integrals. Typical grid 
sizes used are of order $0.5-2.5\times10^6$ points with energy step
of order $10^{-4}-10^{-3}$~eV.

After convergence we calculate the current according to (\ref{current}) and
(\ref{appGFKeldysh}). We also use the converged results to get the 
nonequilibrium electronic density of states
\begin{equation}
 \label{dos}
 A(E) = i\left(G^{>}(E)-G^{<}(E)\right)
\end{equation}
and the nonequilibrium electronic distribution in the junction
\begin{equation}
 \label{fneq}
 f(E) = \mbox{Im}\left[G^{<}(E)\right]/A(E)
\end{equation}

\section{\label{numerics}Results and Discussion}
Here we present numerical results and compare them to those available in
the literature. We choose the bands in both contacts to be the same, with
half width of $W_K^{(0)}=10$~eV positioned in such way that 
the shifted electronic level is in the middle of the band 
$\bar\varepsilon_0=E_K^{(0)}$ ($K=L,R$). The band width is taken big enough
so that results of the calculation do not depend on this choice.
The tolerance for the self-consistent procedure was $10^{-6}$. 

\begin{figure}[htbp]
\centering\includegraphics[width=\columnwidth]{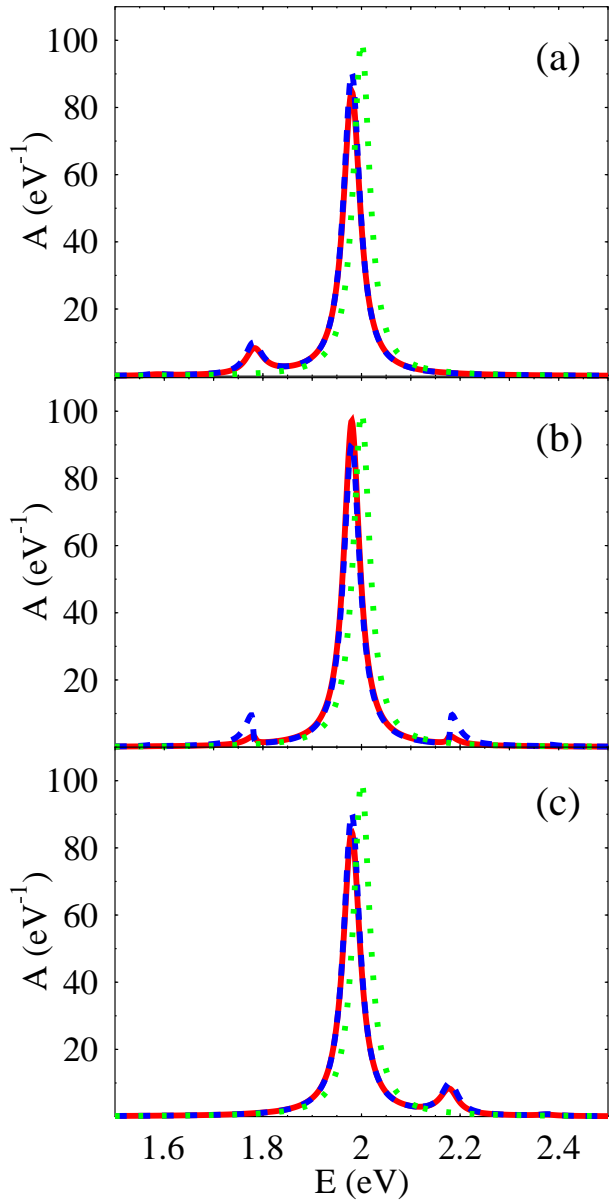}
\caption{\label{fig1}Equilibrium DOS for relatively weak electron-phonon 
coupling: self-consistent result (solid line), zero-order result (dashed line),
uncoupled electron (dotted line). Shown are case of filled (a),
partially filled (b), and empty (c) electron level. See text for parameters.
}
\end{figure}

We start by presenting results for the equilibrium density of states $A(E)$.
Figure~\ref{fig1} shows results for relatively weak electron-phonon interaction
$M_a\sim\Gamma$. The parameters used in this calculation are 
$T=10$~K, $\Gamma_K^{(0)}=0.02$~eV ($K=L,R$),
$\varepsilon_0=2$~eV, $\omega_0=0.2$~eV, $M_a=0.063$~eV, $\gamma_{ph}=0.001$~eV.
This choice corresponds to reorganization energy of $\sim 0.02$~eV, 
Eq.(\ref{bare0}).
Results are presented for several choices of Fermi energy position:
$E_F-\bar\varepsilon_0=2$~eV (a), $0$~eV (b), and $-2$~eV (c).
These cases correspond to filled, partially filled and empty (hole conduction,
mixed conduction and electron conduction respectively). 
The solid line represents
the result of a full self-consistent calculation, the dashed line shows 
the result obtained in the first iteration step (zero-order), 
and the dotted line shows the DOS $A_0$ in the
absence of coupling to phonons. First, in all the pictures 
the polaron shift of the central (elastic) peak position relative to that
of $A_0$ is evident. Second, due to the low temperature and the relatively weak
electron-phonon coupling only one phonon emission peak appears in the plot.  
This peak is symmetric relative to the electron level position (central peak) 
for hole and electron transport (compare Fig.~\ref{fig1}a and c). 
In the intermediate regime, Fig.~\ref{fig1}b, both satellites appear.
In this modest range of electron-phonon interaction the zero-order and 
the self-consistent results are very close. Note that the zero-order result
for filled (Fig.~\ref{fig1}a) and empty (Fig.~\ref{fig1}c) levels
corresponds to single particle transport. In particular, the dashed line in
Fig.~\ref{fig1}c is exactly the scattering theory result presented in
Ref.~\onlinecite{Wingreen} (see Fig.2a there). Note also that,
at least in the low temperature regime, both the scattering
theory approach of \cite{Wingreen} and zero-order approaches within NEGF
in the way it is implemented in \cite{Lundin,Balatsky} 
do not describe correctly the hole part of single particle transport.

\begin{figure}[htbp]
\centering\includegraphics[width=\columnwidth]{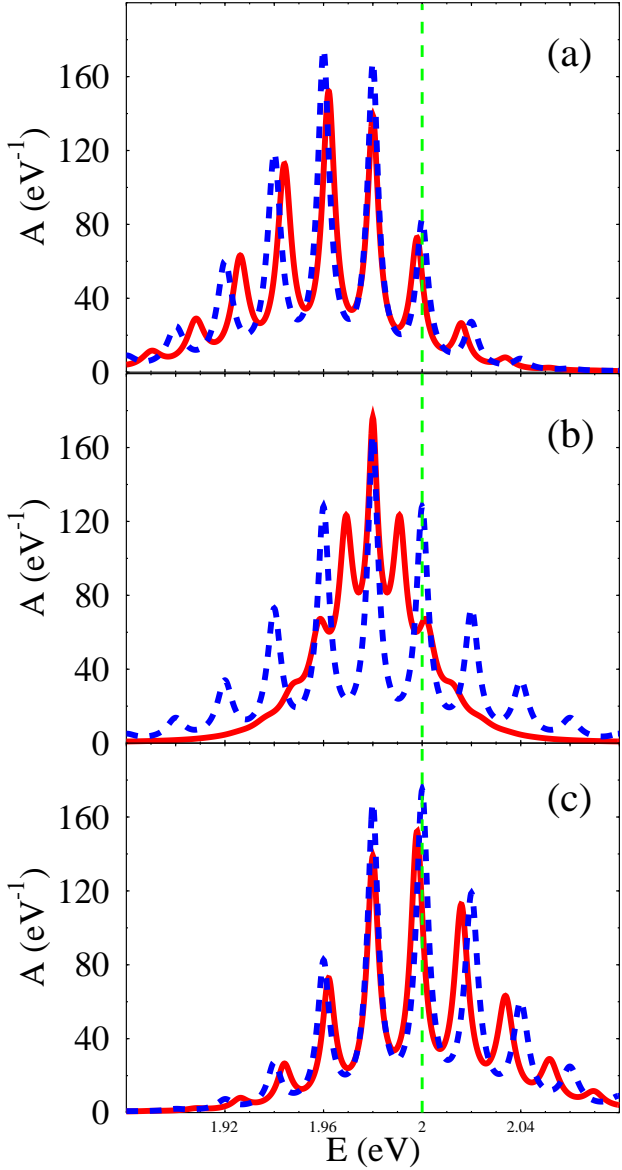}
\caption{\label{fig2}Equilibrium DOS for relatively strong electron-phonon
coupling: self-consistent result (solid line) and zero-order result 
(dashed line). Shown are case of filled (a),
partially filled (b), and empty (c) electron level. See text for parameters.
Dashed vertical line indicates the position of the DOS peak in the absence of 
coupling to phonons. 
}
\end{figure}

Phonon absorption and emission sidebands are seen at higher temperature 
and weaker coupling between bridge and leads (Figure~\ref{fig2}). 
The parameters of this calculation are 
$T=300$~K, $\Gamma_K^{(0)}=0.002$~eV ($K=L,R$),
$\varepsilon_0=2$~eV, $\omega_0=0.02$~eV, $M_a=0.02$~eV, $\gamma_{ph}=0.001$~eV.
The choice corresponds to the same reorganization energy of $\sim 0.02$~eV.
However the electron-phonon coupling here is much more pronounced
due to weaker coupling to the contacts (which implies that the electron-phonon
interaction is relatively much stronger in this case). 
Once more Fig.~\ref{fig2}a is mirror
symmetric of Fig.~\ref{fig2}c and a polaronic shift of $0.02$~eV between
elastic peak in $A$ and $A_0$ is observed. 
The stronger effective electron-phonon coupling is manifested in the 
appearance of five emission phonon sidebands and three peaks corresponding 
to absorption and in the fact that the difference between zero-order
and self-consistent results is much more pronounced here. Indeed,
renormalization due to the interplay between the electron-phonon interaction
and the electronic populations in the leads (the Fermi distribution
in the contacts influences the phonons which in turn affect the electron 
energy distribution on the bridge) may change peak heights and positions 
(see Fig.~\ref{fig2}a and c) or even influence the DOS shape drastically 
(see Fig.~\ref{fig2}b). 
This effect is referred to in Ref.~\onlinecite{Mitra} as ``floating'' of the phonon
sidebands. The zero-order (dashed) curves in Fig.~\ref{fig2}b and c 
can be compared to the corresponding curves in Fig.~7 of Ref.~\onlinecite{Kral}. 
Note that our scheme
in zero-order is similar to the NLCE approach of \cite{Kral} (both 
approaches utilize a cumulant expansion). 
However, while the NLCE appears to be
problematic when going to the second cluster approximation, our scheme
is rather stable when higher order correlations are included by 
the self-consistent procedure. Moreover, we take the mutual influence of 
the electron and phonon subsystems
into account rather than assuming phonons in thermal equilibrium, as is 
done in other work~\cite{Lundin,Balatsky,Bratkovsky,Kral}.
As was already mentioned, this self-consistent aspect of the calculation 
may change the results substantially.

\begin{figure}[htbp]
\centering\includegraphics[width=1.1\columnwidth]{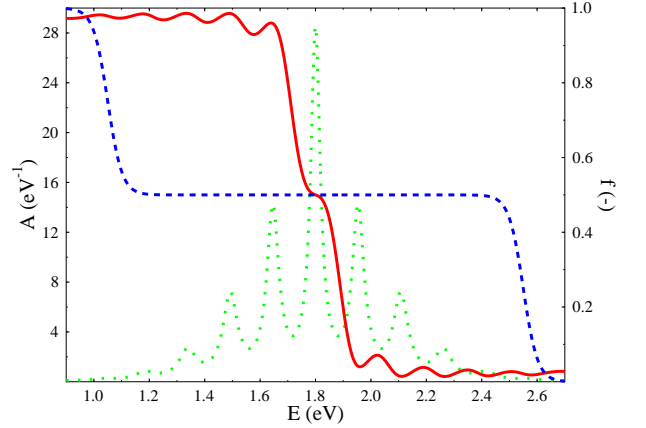}
\caption{\label{fig3}Self-consistent calculation of non-equilibrium DOS 
(dotted line, left vertical axis) and non-equilibrium electron energy distribution 
(solid line, right vertical axis). Shown also is the energy distribution for the 
uncoupled electron (dashed line). See text for parameters.
}
\end{figure}

Figure~\ref{fig3} presents the nonequilibrium DOS (dotted line) 
and the distribution function (solid line) obtained from the
self-consistent calculation. The distribution function in the junction
in the absence of coupling to the phonon (dashed line) is shown for comparison.
Parameters of the calculation are $T=300$~K, $\Gamma_K^{(0)}=0.02$~eV ($K=L,R$),
$\varepsilon_0=2$~eV, $\omega_0=0.2$~eV, $M_a=0.2$~eV, $\gamma_{ph}=0.01$~eV
(corresponds to reorganization energy of $\sim 0.2$~eV). Voltage drop 
is $\Phi=1.5$~V with $E_F=1.8$~eV and $\mu_K=E_F\pm\Phi/2$ ($K=L,R$).
This setup corresponds to a mixed (both electron and hole) transport through 
the junction (analog to graph b in Figs.~\ref{fig1} and \ref{fig2}).
Local minima in the nonequilibrium population $f$ to the left of 
the elastic peak
(local maxima to the right) correspond to positions of phonon sidebands
in the DOS $A$. The effect is due to outscattering of electrons from the
energy regions (outscattering of holes from or equivalently inscattering of 
electrons into the energy regions) due to phonon emission. One sees that
strong electron-phonon interaction essentially changes the distribution 
function. Similar results were reported in Ref.~\onlinecite{Kral} (see Fig.6 there).

\begin{figure}[htbp]
\centering\includegraphics[width=1.1\columnwidth]{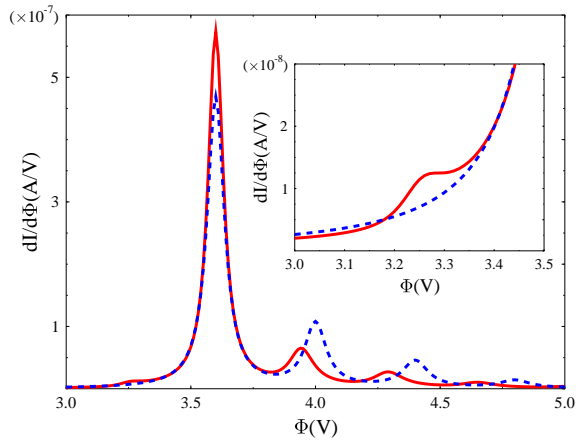}
\caption{\label{fig4}Differential conductance vs. source-drain voltage.
Shown are self-consistent (solid line) and zero-order (dashed line) results.
Enlarged phonon absorption peak in the self-consistent result is reproduced 
in the inset. See text for details.
}
\end{figure}

Figure~\ref{fig4} presents self-consistent (solid line) and zero-order 
(dashed line) results for the differential conductance $dI/d\Phi$ as a function 
of the applied source-drain voltage $\Phi$.
Parameters of the calculation are the same as in Fig.~\ref{fig3} except
that the calculation is done at low temperature $T=10$~K.
Phonon assisted resonant tunneling reveals itself in conductance peaks
associated with different vibronic resonances, i.e. electronic levels
dressed by different numbers of phonon excitations.
As was mentioned above,
renormalization due to electron-phonon interaction leads to shift
of position and height of phonon sideband peaks.
The zero-order (dashed) curve can be compared to Fig.~6 of Ref.~\onlinecite{Lundin}.
A surprising feature is formation of an additional peak at $\Phi\sim 3.25$~V 
(see inset) in the self-consistent (solid) curve. 
While the low temperature of the calculation
makes phonon absorption unlikely, we suspect that the effect may be
a result of phonon absorption due to heating of the phonon subsystem by
electron flux. Clearly such a scenario is possible only if
coupling to the contacts is treated beyond second order, 
i.e. within our scheme only the self-consistent calculation
can yield the effect.

\begin{figure}[htbp]
\centering\includegraphics[width=1.1\columnwidth]{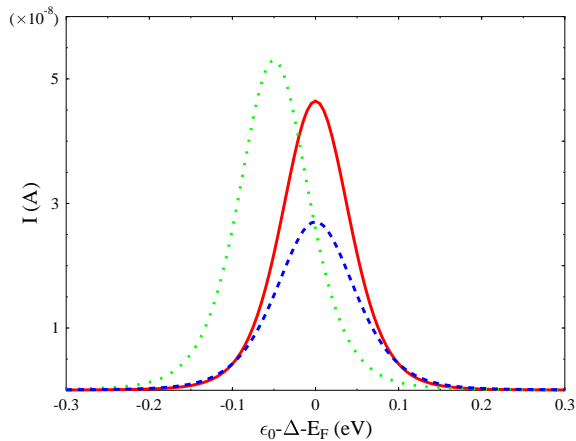}
\caption{\label{fig5}Current vs. gate potential
when source-drain voltage is fixed at $\Phi=0.01$~V. 
The position of the shifted level is given relative to the
unbiased Fermi energy.
Shown are the self-consistent (solid line), zero-order 
(dashed line), and uncoupled electron (dotted line) results.
See text for details.
}
\end{figure}

Figure~\ref{fig5} presents current plotted against the energy of the bridge 
electronic level (that can be controlled by a gate voltage) at small fixed 
source-drain voltage. The horizontal axis represents
the position of the shifted electron level relative to the original (unbiased) 
Fermi energy (average chemical potential of the contacts).
Parameters of the calculation are $T=10$~K, $\Gamma_K^{(0)}=0.005$~eV ($K=L,R$),
$\omega_0=0.05$~eV, $M_a=0.05$~eV, $\gamma_{ph}=0.001$~eV
(corresponds to reorganization energy of $\sim 0.05$~eV). The source-drain
voltage drop is $\Phi=0.01$~V. Shown are self-consistent (solid line) and
zero-order (dashed line) results. Also shown is current profile when
electron and phonon are decoupled (dotted line). 
The shift in the peak position is due to the reorganization energy. 

\begin{figure}[htbp]
%\vspace*{-3cm}
\centering\includegraphics[angle=-90,width=\columnwidth]{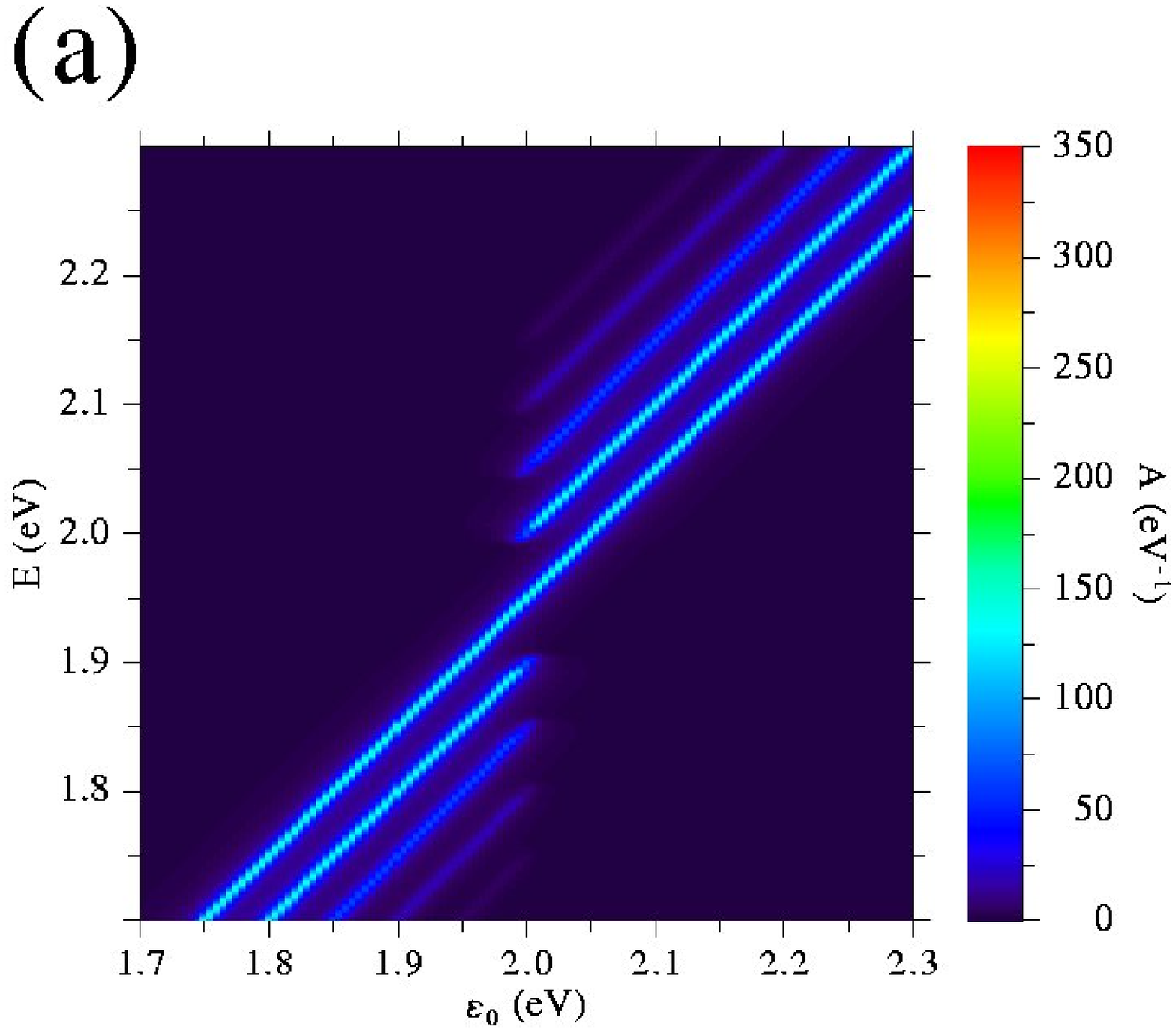}\\*[-1.5cm]
\vspace*{0.5cm}
\centering\includegraphics[angle=-90,width=\columnwidth]{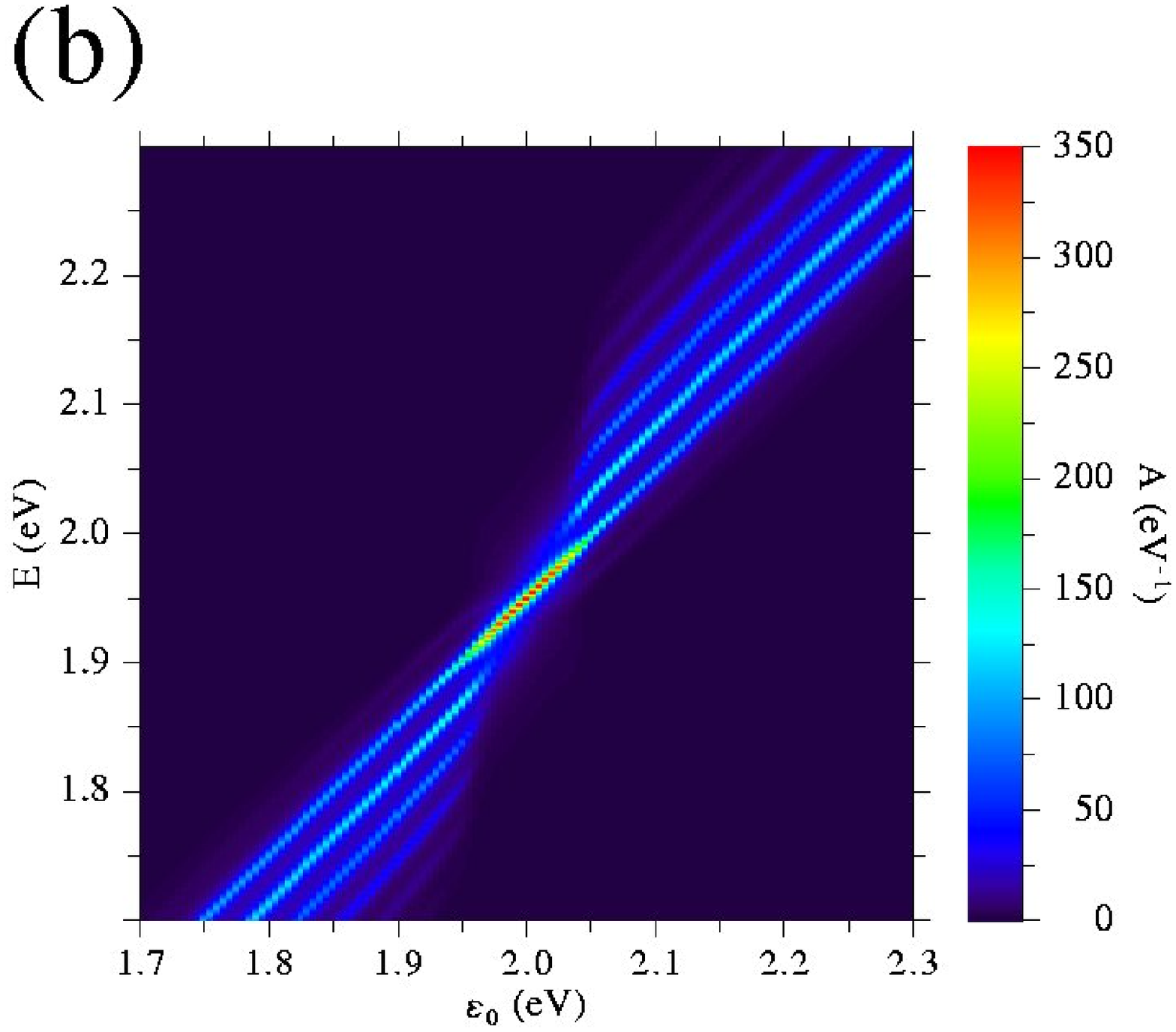}
\caption{\label{fig6}Contour plot of non-equilibrium DOS vs. energy and
position of electron level: zero-order (a) and self-consistent (b) results.
Parameters of calculation are the same as in Fig.~\ref{fig5}.
}
\end{figure}

Mitra~et~al.~\cite{Mitra} have found that for small source-drain voltage
($\Phi=\Phi_{source}-\Phi_{drain}<\omega_0$) no phonon sideband appears
in the plot of source-drain current vs. gate voltage potential.
This contradicts earlier findings (see Refs.~\onlinecite{Wingreen,Lundin,Balatsky})
and since consideration in Ref.~\onlinecite{Mitra} is based on 
the Migdal-Eliashberg theory, which is known to break down in the case of 
intermediate to strong electron-phonon coupling~\cite{Mahan,Hague,Schonhammer}, 
this finding should be critically examined. The results of Fig.~\ref{fig5}
confirm the conclusions of Mitra~et~al.
Note also that the cause of the previous erroneous predictions
is neglect or oversimplified description of hole transport~\cite{FCeh} in 
\cite{Wingreen,Lundin,Balatsky}. 
Finally, in the case where the couplings to the two contacts are 
proportional to each other, $\Gamma_L(E)=x\Gamma_R(E)$, the current through 
the junction for resonant level model can be obtained as an integral over 
DOS~\cite{HaugJauho,current},
\begin{equation}
 \label{IA}
 I_{s-d} = \frac{e}{\hbar}\int\frac{dE}{2\pi}\,
 \frac{\Gamma_L(E)\Gamma_R(E)}{\Gamma(E)} A(E)
 \left[f_L(E)-f_R(E)\right]
\end{equation} 
In the case of small source-drain voltage, where $\Gamma(E)$ is a smooth
function of $E$, the current characteristics are determined by $A(E)$.
We can therefore demonstrate the ``floating'' behavior of phonon sidebands in 
a contour plot of the DOS vs. energy and electron level position 
(see Fig.~\ref{fig6}). Parameters of the calculation here are the same 
as in Fig.~\ref{fig5}. Figure~\ref{fig6}a shows the zero-order result, while
Figure~\ref{fig6}b presents a self-consistent calculation.
One sees that the phonon sidebands disappear around the position of Fermi level
$E=E_F=1.95$~eV. This in turn leads to absence of peaks in the 
current lineshape of Fig.~\ref{fig5}. Note that since the effect is
present already in the zero-order situation (Fig.~\ref{fig6}a),
it can be studied analytically. We find, that for low temperatures
($T\to 0$) one can not see phonon sidebands in the current-voltage
characteristic while changing gate voltage unless $\Phi>\omega_0$
(see Appendix~\ref{appE}).

\section{\label{conclude}Conclusion}
Within a non-equilibrium Green function formalism on the Keldysh contour, 
we have developed an approximate self-consistent procedure for 
treating a phonon-assisted
resonant level model in the case of intermediate to strong 
electron-phonon interaction, where the strength of this interaction is
determined relative to the bridge-contacts coupling. 
Our scheme goes beyond earlier considerations of this problem
by taking into account the mutual influence of the electron and phonon
subsystems. In zero-order and in a single electron transport situation 
(Fermi energies in the leads far above or far below the bridge level
so that the latter is full or empty, respectively)
it is similar to other approaches~\cite{Wingreen,Lundin,Balatsky,Bratkovsky}.
However in the case of a partially filled resonance level even the zero-order
of the self-consistent scheme extends previous calculations 
(at least in low temperature regime) in
treating correctly hole transport. Our approach is also similar in zero-order
to the NLCE scheme proposed in~\cite{Kral}, since both use cumulant expansion
on the contour. However while NLCE appears to be
problematic when going to the second cluster approximation, the present scheme
is rather stable when higher order correlations 
are included by the self-consistent procedure. 

We have presented several numerical examples and compared results to those 
of earlier studies. The self-consistent calculation is found to yield
drastically different results as compared to the zero-order
theory in the case of strong electron-phonon interaction and in the 
region of a partially filled electronic level. In particular, it leads 
to shifts (in position and height) of peaks in the conductance vs. 
source-drain voltage plot and to phonon absorption signals even at low
temperatures, that probably result from heating of the primary phonon by the
electronic flux.
The non-equilibrium electron DOS and the electron distribution show
a similar structure, with peaks associated with phonon emission and 
absorption.
Finally, we confirm the statement of Ref.~\onlinecite{Mitra} that current measured 
in a gate voltage experiment, when source-drain voltage is fixed at some small 
value, will not produce peaks in current vs. position of electron level plot,
as was erroneously suggested in~\cite{Wingreen,Lundin,Balatsky}.
Since the statement in~\cite{Mitra} is based on application of the 
Migdal-Eliashberg theory, known to break down in the case of intermediate 
to strong electron-phonon interaction, our confirmation seems to be essential.
Together with previous work~\cite{IETS_SCBA}, which deals with the
weak electron-phonon interaction case in a self-consistent manner,
this provides tools for describing both resonant 
(intermediate to strong interaction) 
and off-resonant (weak interaction) tunneling regimes in molecular 
junctions. 

Straightforward generalization of the present scheme would involve
retaining mixed terms in deriving the equation of motion for 
the phonon Green function (Appendix~\ref{appC}),
thus abandoning the non-crossing approximation 
(i.e. introducing vertex corrections). One could
also go beyond second order in electron-phonon coupling in the cumulant 
expansion (Appendix~\ref{appB}) or in coupling to the contacts in EOMs 
(Appendices~\ref{appC} and \ref{appD}). 
Development of a scheme spanning the entire range of 
parameters in the nonequilibrium situation is a goal for future work.

\begin{acknowledgments}
We are grateful to the NASA URETI program, the NSF-NCN program through
Purdue University and the MolApps program of DARPA for support.
The research of AN is supported by the Israel Science Foundation and by the
US-Israel Binational Science Foundation.
\end{acknowledgments}

\appendix

\section{\label{appA}Hamiltonian transformation}
% Derivation of Eq.(\ref{barH})
Here we derive Eq.(\ref{barH}) for the case of weak coupling between 
the primary phonon and the thermal bath (see below). 
Let us focus on the part of the Hamiltonian~(\ref{H}) relevant to the
electron--phonon interaction
\begin{eqnarray}
 \label{Hel_ph}
 \hat H_{el-ph} &=& \varepsilon_0\hat c^\dagger\hat c +
 \omega_0\hat a^\dagger\hat a +
 \sum_\beta\omega_\beta\hat b^\dagger_\beta\hat b_\beta
 \nonumber \\ &+&
 M_a\hat Q_a\hat c^\dagger\hat c +
 \sum_\beta U_\beta\hat Q_a\hat Q_\beta
\end{eqnarray}
We use the small polaron transformation, Eqs.~(\ref{small_polaron})
and (\ref{Sa}), to get
\begin{align}
 &e^{\hat S_a}\hat H_{el-ph} e^{-\hat S_a} =
 \left(\varepsilon_0-\frac{M_a^2}{\omega_0}\right)\hat c^\dagger\hat c +
 \omega_0\hat a^\dagger\hat a +
 \sum_\beta\omega_\beta\hat b^\dagger_\beta\hat b_\beta
 \nonumber \\ &\quad +
 \sum_\beta U_\beta\hat Q_a\hat Q_\beta
 - 2\sum_\beta M_a\frac{U_\beta}{\omega_\beta}\hat Q_\beta\hat c^\dagger\hat c
\end{align}
One more transformation 
\begin{equation}
 \label{small_polaron_b}
 e^{\hat S_a}\hat H_{el-ph} e^{-\hat S_a} \to 
 e^{\hat S_b}e^{\hat S_a}\hat H_{el-ph} e^{-\hat S_a} e^{-\hat S_b}
\end{equation}
with
\begin{equation}
 \label{Sb}
 \hat S_b = -2 M_a\sum_\beta\frac{U_\beta}{\omega_0\omega_\beta}
 \left(\hat b^\dagger_\beta-\hat b_\beta\right) \hat c^\dagger\hat c
\end{equation}
will lead to Hamiltonian $\hat H_{el-ph}^{(1)}$ of the form (\ref{Hel_ph})
with substitutions
\begin{equation}
 \varepsilon_0 \to \varepsilon_0 - \Delta^{(1)} 
 \qquad \mbox{and} \qquad M_a \to M_a^{(1)}
\end{equation}
where renormalized parameters are
\begin{eqnarray}
 \Delta^{(1)} &=& \frac{M_a^2}{\omega_0}
 \left(1+\sum_\beta\frac{(2U_\beta)^2}{\omega_0\omega_\beta}\right)
 \\
 M_a^{(1)} &=& M_a\sum_\beta\frac{(2U_\beta)^2}{\omega_0\omega_\beta}
\end{eqnarray}
Repeating these transformations a second time leads to a Hamiltonian 
$\hat H_{el-ph}^{(2)}$ of the same form but with different parameters
\begin{align}
 &\Delta^{(2)} =
 \Delta^{(1)}+\frac{\left(M_a^{(1)}\right)^2}{\omega_0}
 \left(1+\sum_\beta\frac{(2U_\beta)^2}{\omega_0\omega_\beta}\right)
 \\
 &= \frac{M_a^2}{\omega_0}\left(1+\sum_\beta\left\{
 \frac{(2U_\beta)^2}{\omega_0\omega_\beta}
 +\left[\frac{(2U_\beta)^2}{\omega_0\omega_\beta}\right]^2
 +\left[\frac{(2U_\beta)^2}{\omega_0\omega_\beta}\right]^3
 \right\}\right)
 \nonumber \\
 &M_a^{(2)} = M_a^{(1)}\sum_\beta\frac{(2U_\beta)^2}{\omega_0\omega_\beta}
 = M_a\left[\sum_\beta\frac{(2U_\beta)^2}{\omega_0\omega_\beta}\right]^2
\end{align}
Repeating the procedure described above, we get for the $n^{th}$ step
\begin{eqnarray}
 \Delta^{(n)} &=& \frac{M_a^2}{\omega_0}\sum_{i=0}^{2n-1}
 \left[\sum_\beta\frac{(2U_\beta)^2}{\omega_0\omega_\beta}\right]^i
 \\
 M_a^{(n)} &=& 
 M_a\left[\sum_\beta\frac{(2U_\beta)^2}{\omega_0\omega_\beta}\right]^n
\end{eqnarray}
Now, assuming that coupling between primary phonon and thermal bath is 
small in the sense
\begin{equation}
 \sum_\beta\frac{(2U_\beta)^2}{\omega_0\omega_\beta} < 1
\end{equation}
then continuing the procedure in the limit $n\to\infty$ leads to
\begin{eqnarray}
 \Delta^{(\infty)} &=& \frac{M_a^2}{\omega_0}
 \frac{1}{1-\sum_\beta (2U_\beta)^2/\omega_0\omega_\beta}
 \\
 M_a^{(\infty)} &=& 0
\end{eqnarray}
So we arrive at decoupling of electron and phonon degrees of freedom in
$\hat H_{el-ph}^{(\infty)}$. Going back to the full Hamiltonian~(\ref{H})
of the system we note that the true complete separation of the 
electronic and phononic degrees of freedom, as achieved in 
$\hat H_{el-ph}^{(\infty)}$, is impossible here due to the coupling 
between the resonant level and the leads.  
Instead the aforementioned procedure leads to
\begin{align}
 \label{Hinfty}
 \hat H^{(\infty)} &= (\varepsilon_0-\Delta^{(\infty)})\hat c^\dagger\hat c +
 \sum_{k\in\{L,R\}} \varepsilon_k \hat c_k^\dagger\hat c_k
 \nonumber \\ &+
 \sum_{k\in\{L,R\}} \left(V_k\hat c_k^\dagger\hat c\hat X^{(\infty)} 
                         + \mbox{H.c.}\right)
 \\ &+
 \omega_0\hat a^\dagger\hat a + 
 \sum_\beta\omega_\beta\hat b^\dagger_\beta\hat b_\beta +
 \sum_\beta U_\beta\hat Q_a\hat Q_\beta
 \nonumber
\end{align}
where
\begin{align}
 \hat X^{(\infty)} &= \hat X_a^{(\infty)} \hat X_b^{(\infty)}
 \\
 \hat X_a^{(\infty)} &= 
 \exp\left[-\frac{M_a}{\omega_0(1-\sum_\beta(2U_\beta)^2/\omega_0\omega_\beta)}
 \left(\hat a^\dagger-\hat a\right)\right]
 \\
 \hat X_b^{(\infty)} &= \prod_\beta\exp\left[
 \frac{M\,2U_\beta}
 {\omega_0\omega_\beta(1-\sum_\beta(2U_\beta)^2/\omega_0\omega_\beta)}
 \right.\nonumber\\&\qquad\qquad\times\left.
 \left(\hat b^\dagger_\beta-\hat b_\beta\right)\right]
\end{align}
Finally, neglecting in the spirit of the non-crossing approximation the
$\hat X_b^{(\infty)}$ operators in the coupling to the contacts, 
renormalizing $M_a$ to incorporate the denominator
in the exponent of $\hat X_a$ expression, and setting 
$\Delta=\Delta^{(\infty)}$, leads to (\ref{barH}).

\section{\label{appB}$<T_c \hat X(\tau_1)\hat X^\dagger(\tau_2)>$ 
in terms of $D_{P_aP_a(\tau_1,\tau_2)}$}
% <XX> in terms of D
Here we derive Eq.(\ref{XXKeldysh}) expressing the shift generator
correlation function in terms of the phonon momentum Green function.
We start from the Taylor series that expresses the correlation function 
as a sum of moments
\begin{align}
 \label{XXmoment}
 <T_c \hat X(\tau_1)\hat X^\dagger(\tau_2)> &= 
 \sum_{n=0}^\infty \sum_{m=0}^\infty 
 \frac{(-i\lambda_a)^n}{n!} \frac{(i\lambda_a)^m}{m!} 
 \nonumber \\ &\times
 <T_c \hat {P_a}^n(\tau_1) \hat {P_a}^m(\tau_2)>
\end{align}
where Eq.~(\ref{Xa}) was used. The cumulant expansion for this function is
\begin{align}
 \label{XXcumulant}
 &<T_c \hat X(\tau_1)\hat X^\dagger(\tau_2)> =
 \exp\left[\sum_{p=1}^\infty\frac{\lambda_a^p}{p!}
           \varphi_p(\tau_1,\tau_2)\right]
 \nonumber \\
 &\quad =
 1 + \sum_{p=1}^\infty \frac{\lambda_a^p}{p!}\varphi_p(\tau_1,\tau_2)
 \\ &\quad +
 +  \frac{1}{2}\sum_{p_1=1}^\infty\sum_{p_2=1}^\infty
   \frac{\lambda_a^{p_1}}{p_1!} \frac{\lambda_a^{p_2}}{p_2!}
   \varphi_{p_1}(\tau_1,\tau_2)\varphi_{p_2}(\tau_1,\tau_2)
 + \ldots
 \nonumber
\end{align}
where $\varphi_p(\tau_1,\tau_2)$ is a cumulant of order $p$.
Considering terms up to $\lambda_a^2$ and
equating same orders of $\lambda_a$ in (\ref{XXmoment}) and (\ref{XXcumulant})
leads to 
\begin{equation}
 \label{XXmomcum}
 \left\{
 \begin{array}{l}
    i<\hat P_a(\tau_1)> - i<\hat P_a(\tau_2)> = \varphi_1(\tau_1,\tau_2) \\
    2<T_c \hat P_a(\tau_1)\hat P_a(\tau_2)> - <\hat {P_a}^2(\tau_1)>
    - <\hat {P_a}^2(\tau_2)> \\ \quad = 
    \varphi_2(\tau_1,\tau_2) + {\varphi_1}^2(\tau_1,\tau_2)
 \end{array}
 \right.
\end{equation}
In steady-state 
$<\hat {P_a}^n(\tau_1)> = <\hat {P_a}^n(\tau_2)> = <\hat {P_a}^n>$,
which implies 
\begin{equation}
 \label{XXmomcumSS}
 \left\{
 \begin{array}{l}
    \varphi_1(\tau_1,\tau_2)=0 \\
    \varphi_2(\tau_1,\tau_2)= 2<T_c \hat P_a(\tau_1)\hat P_a(\tau_2)>
    - 2<\hat {P_a}^2>
 \end{array}
 \right.
\end{equation}
Using this in (\ref{XXcumulant}) leads to (\ref{XXKeldysh}).

\section{\label{appC}EOM for $D_{P_aP_a}(\tau_1,\tau_2)$}
% EOM for D
Here we derive a Dyson-like equation for the phonon momentum Green function
$D_{P_aP_a}$ under the Hamiltonian (\ref{barH}) with the EOM method.
Under Hamiltonian (\ref{barH}), the EOM for the shift and momentum operators,
Eqs.~(\ref{Q}) and (\ref{P}), in the Heisenberg picture on the Keldysh contour
are
\begin{eqnarray}
 \label{EOMPa}
 i\frac{\partial\hat P_a(\tau)}{\partial\tau} &=&
 -i\omega_0\hat Q_a(\tau) - 2i\sum_\beta U_\beta \hat Q_\beta(\tau)
 \\
 \label{EOMQa}
 i\frac{\partial\hat Q_a(\tau)}{\partial\tau} &=&
 i\omega_0\hat P_a(\tau) 
 \\ &-& 2\lambda_a\sum_{k\in\{L,R\}}\left(
 V_k\hat c_k^\dagger(\tau)\hat c(\tau)\hat X_a(\tau)-\mbox{H.c}\right)
 \nonumber \\
 \label{EOMPbeta}
 i\frac{\partial\hat P_\beta(\tau)}{\partial\tau} &=&
 -i\omega_\beta\hat Q_\beta(\tau) - 2iU_\beta \hat Q_a(\tau)
 \\
 \label{EOMQbeta}
 i\frac{\partial\hat Q_\beta(\tau)}{\partial\tau} &=&
 i\omega_\beta\hat P_\beta(\tau)
\end{eqnarray} 
We are looking for a Dyson-like equation for the phonon momentum Green function
(\ref{D}). We introduce the operator
\begin{equation}
 \label{invD0}
 \hat D_{P_aP_a}^{(0)}{}^{-1} = -\frac{1}{2\omega_0}
 \left[\frac{\partial^2}{\partial\tau^2}+\omega_0^2\right]
\end{equation}
with property
\begin{equation}
 \label{EOMD0}
 \hat D_{P_aP_a}^{(0)}{}^{-1} \cdot D_{P_aP_a}^{(0)}(\tau,\tau') =
 D_{P_aP_a}^{(0)}(\tau,\tau') \cdot \hat D_{P_aP_a}^{(0)}{}^{-1} =
 \delta(\tau,\tau')
\end{equation}
where $D_{P_aP_a}^{(0)}$ is the Green function of a free phonon 
(decoupled both from the electron and the thermal bath).
Applying (\ref{invD0}) to (\ref{D}) from the left,
taking into account (\ref{EOMPa})-(\ref{EOMQbeta}), and
restricting consideration to the non-crossing approximation (NCA)\cite{Bickers},
so that terms mixing different processes are disregarded, one gets
\begin{align}
 \label{EOMDstep1}
 &\hat D_{P_aP_a}^{(0)}{}^{-1} D_{P_aP_a}(\tau,\tau') = 
 \delta(\tau,\tau') +
 \sum_\beta U_\beta\frac{\omega_\beta}{\omega_0}D_{P_\beta P_a}(\tau,\tau')
 \nonumber \\ &+
 i\lambda_a\sum_{k\in\{L,R\}}\left[
 V_k(-i)<T_c\hat c_k^\dagger(\tau)\hat c(\tau)\hat X_a(\tau)\hat P_a(\tau')
 \right. \\ & \left.\quad\qquad
 -V_k^{*}(-i)<T_c\hat c^\dagger(\tau)\hat X_a^\dagger(\tau)\hat c_k(\tau)
 \hat P_a(\tau')>\right]
 \nonumber
\end{align}
Next we apply operator (\ref{invD0}) to (\ref{EOMDstep1}) from the right.
The procedure is the same in the sense that here we deal once more with
EOM for $\hat P_a$ (the one depending on $\tau'$).
After tedious but straightforward algebra and convolution of the result
with $D_{P_aP_a}^{(0)}$ we obtain a Dyson-like equation 
\begin{align}
 \label{likeDysonD}
 &D_{P_aP_a}(\tau,\tau') = D_{P_aP_a}^{(0)}(\tau,\tau')
 \\
 &\quad + \int_c d\tau_1 \int_c d\tau_2\, D_{P_aP_a}^{(0)}(\tau,\tau_1)\,
 \Pi_{P_aP_a}(\tau_1,\tau_2)\, D_{P_aP_a}^{(0)}(\tau_2,\tau')
 \nonumber
\end{align}
with the self-energy expression
\begin{align}
 \label{likeSED}
 &\Pi_{P_aP_a}(\tau_1,\tau_2) = 
 \sum_\beta\frac{2U_\beta^2\omega_\beta}{\omega_0^2}\delta(\tau_1,\tau_2) 
 \\
 & +
 \sum_\beta\left(U_\beta\frac{\omega_\beta}{\omega_0}\right)^2
 D_{P_\beta P_\beta}(\tau_1,\tau_2)
 -i\lambda_a^2\sum_{k\in\{L,R\}}|V_k|^2
 \nonumber \\ & \times
 \left[g_k(\tau_2,\tau_1)
 G_c^{(0)}(\tau_1,\tau_2)<T_c\hat X_a(\tau_1)\hat X_a^\dagger(\tau_2)>_0
 +\mbox{H.c.} \right]
 \nonumber
\end{align}
In what follows we neglect renormalization of the phonon frequency due to 
coupling to the thermal bath (first term on the right and real part of the 
second term in the self-energy expression). This is in analogy to the wide band
approximation (for discussion on applicability of the approximation to 
coupling to the thermal bath case see Ref.~\onlinecite{IETS_SCBA}). 
In the second term on 
the right we also replace $\omega_\beta/\omega_0$ by unity arguing that main
contribution to $D_{P_aP_a}$ comes from the region $\omega_\beta\sim\omega_0$.
Then the dressed form (zero-order Green and correlation functions are 
substituted by the full ones) of (\ref{likeSED})  is Eq.(\ref{DSEKeldysh}).

\section{\label{appD}EOM for $G_c(\tau_1,\tau_2)$}
% EOM for G_c 
Here we derive a Dyson-like equation for electron Green function $G_c$
under Hamiltonian (\ref{barH}) using the EOM method. 
First we note that under Hamiltonian (\ref{barH}), the EOM for 
operator $\hat c$ in the Heisenberg picture on the Keldysh contour is
\begin{equation}
 \label{EOMc}
 i\frac{\partial\hat c(\tau)}{\partial\tau} = 
 \bar\varepsilon_0\hat c(\tau) + \sum_{k\in\{L,R\}} V_k^{*}
 \hat X_a^\dagger(\tau)\hat c_k(\tau)
\end{equation}
with the corresponding Hermitian conjugate for the creation operator 
$\hat c^\dagger$. We then introduce the operator
\begin{equation}
 \label{invG0}
 \hat G_c^{(0)}{}^{-1} = i\frac{\partial}{\partial\tau}-\bar\varepsilon_0
\end{equation}
with the property 
\begin{equation}
 \label{EOMGc0}
 \hat G_c^{(0)}{}^{-1} \cdot G_c^{(0)}(\tau,\tau') =
 G_c^{(0)}(\tau,\tau') \cdot \hat G_c^{(0)}{}^{-1} = \delta(\tau,\tau')
\end{equation}
Here $G_c^{(0)}(\tau,\tau')$ is the Green function for an electronic level 
decoupled from the contacts.
Applying it to the electron Green function $G_c(\tau,\tau')$ first from the 
left and then from the right, one gets
\begin{align}
 \label{likeDysonGc}
 &\hat G_c^{(0)}{}^{-1} \cdot G_c(\tau,\tau') \cdot \hat G_c^{(0)}{}^{-1}
 = \delta(\tau,\tau') \cdot \hat G_c^{(0)}{}^{-1}
 \\ &\quad +
   \sum_{k\in\{L,R\}} |V_k|^2 g_k(\tau,\tau') 
   <T_c\hat X_a(\tau')\hat X_a^\dagger(\tau)>
 \nonumber
\end{align}
Finally, we take convolution of (\ref{likeDysonGc}) with $G_c^{(0)}$ from left 
and right and utilize Eq.(\ref{EOMGc0}) to arrive at a Dyson-like equation
for $G_c$ with the dressed self-energy of the form presented 
in (\ref{GcSEKeldysh}).

\section{\label{appE}Phonon sidebands in current-voltage characteristics}
% I_{sd} in gate voltage experiment. T=0 case.
Here we study analytically the possibility of observing phonon sidebands
in current-voltage characteristics when the gate voltage 
applied to the junction is varied. We restrict our consideration to the
zero-order situation (first step of the self-consistent procedure)
and low temperature (we take $T=0$). The zero-order
correlation functions for the shift generator operators 
(in the time domain) are
\begin{align}
 \label{XXd_T0}
 &<\hat X_a(t)\hat X^\dagger_a(0)> 
 \\
 &=e^{-\lambda_a^2(2N_0+1)}
 \exp\{\lambda_a^2[(N_0+1)e^{-i\omega_0 t}+N_0e^{i\omega_0 t}]\}
 \nonumber \\
 &\stackrel{T\to 0}{\longrightarrow}
 e^{-\lambda_a^2} \exp\{\lambda_a^2e^{-i\omega_0 t}\}
 \equiv e^{-\lambda_a^2}\sum_{k=0}^{\infty}\frac{\lambda_a^{2k}}{k!}
 e^{-ik\omega_0 t} 
 \nonumber \\
 \label{XdX_T0}
 &<\hat X^\dagger_a(0)\hat X_a(t)> 
 \\
 &=e^{-\lambda_a^2(2N_0+1)}
 \exp\{\lambda_a^2[N_0e^{-i\omega_0 t}+(N_0+1)e^{i\omega_0 t}]\}
 \nonumber \\
 &\stackrel{T\to 0}{\longrightarrow}
 e^{-\lambda_a^2} \exp\{\lambda_a^2e^{i\omega_0 t}\}
 \equiv e^{-\lambda_a^2}\sum_{k=0}^{\infty}\frac{\lambda_a^{2k}}{k!}
 e^{ik\omega_0 t}
 \nonumber
\end{align}
Within the wide-band approximation, the zero-order lesser and greater electron 
Green functions (in the energy domain) are
\begin{eqnarray}
 \label{Gclt_T0}
 G_c^{<}(E) &=& \frac{i\theta(\mu_L-E)\Gamma_L+i\theta(\mu_R-E)\Gamma_R}
                     {(E-\bar\varepsilon_0)^2+(\Gamma/2)^2}
 \\
 \label{Gcgt_T0}
 G_c^{>}(E) &=& \frac{-i\theta(E-\mu_L)\Gamma_L-i\theta(E-\mu_R)\Gamma_R}
                     {(E-\bar\varepsilon_0)^2+(\Gamma/2)^2}
\end{eqnarray}
Applying (\ref{XXd_T0})-(\ref{Gcgt_T0}) to (\ref{appGFKeldysh}),
using the resulting Green function in (\ref{dos}) and the spectral
function $A$ in (\ref{IA}) leads to an expression for the source-drain 
current in the form (putting $\mu_L-\mu_R=\Phi>0$)
\begin{eqnarray}
 \label{I_T0}
 I_{sd} &=& \frac{e}{\hbar}\frac{\Gamma_L\Gamma_R}{\Gamma}
 e^{-\lambda_a^2}\sum_{k=0}^{\infty}\frac{\lambda_a^{2k}}{k!}
 \nonumber \\
 &\times& \int_{\mu_R}^{\mu_L}\frac{dE}{2\pi}\left[
 \frac{\theta(E-k\omega_0-\mu_R)\Gamma_R}
      {(E-k\omega_0-\bar\varepsilon_0)^2+(\Gamma/2)^2} 
 \right. \\ &&\qquad\qquad\left.
 + \frac{\theta(\mu_L-E-k\omega_0)\Gamma_L}
      {(E+k\omega_0-\bar\varepsilon_0)^2+(\Gamma/2)^2}
 \right]
 \nonumber
\end{eqnarray}
while the gate voltage changes the position of $\bar\varepsilon_0$.
It is obvious that one can not observe phonon sidebands ($k>0$ terms)
in the source-drain current-voltage unless $\Phi>\omega_0$.


\begin{thebibliography}{99}
\bibitem{Dltgtcomment}Note that $D^{<}(t=0)=D^{>}(t=0)$. Both appear
in (\ref{XXlt}) and (\ref{XXgt}) in order to demonstrate the projection
rule: if $A(\tau,\tau')=\exp[B(\tau,\tau')]$ on the contour, then
$A^{<,>}(t,t')=\exp[B^{<,>}(t,t')]$.
\bibitem{FCeh}The Franck-Condon factor for electron and hole transport is
different: $<\hat X_a(t)\hat X_a^\dagger(0)>$ for electrons and 
$<\hat X_a^\dagger(0)\hat X_a(t)>$ for holes. 
Refs.~\onlinecite{Wingreen,Lundin,Balatsky} take it in the same (electron-type)
form for both types of transport, and this is the main cause of consequent 
erroneous predictions.
\end{thebibliography}
\end{document}